# The Random Walk behind Volatility Clustering

Sabiou Inoua

inouasabiou@gmail.com

Dec. 2016

(Preliminary version)

**Abstract**

Financial price changes obey two universal properties: they follow a power law and they tend to be clustered in time. The second regularity, known as volatility clustering, entails some predictability in the price changes: while their sign is uncorrelated in time, their amplitude (or volatility) is long-range correlated. Many models have been proposed to account for these regularities, notably agent-based models; but these models often invoke relatively complicated mechanisms. This paper identifies a basic reason behind volatility clustering: the impact of exogenous news on expectations. Indeed the expectations of financial agents clearly vary with the advent of news; the simplest way of modeling this idea is to assume the expectations follow a random walk. We show that this random walk implies volatility clustering in a generic way.

## 1. Background

Financial price changes obey two universal properties: they are distributed according to a power law, with an almost cubic exponent, and they tend to be clustered in time. The first regularity implies that extreme price changes are much more likely than would suggest the normal distribution, for instance. The second property, known as *volatility clustering*, implies that high-amplitude price changes tend to be followed by high-amplitude price changes, and low-amplitude price changes, by low-amplitude price changes. This entails a nontrivial predictability in price changes: while their sign is uncorrelated, its amplitude (or volatility) is long-range dependent.

In formal terms, let $P_t$ be the price of a financial asset at the closing of period $t$, let its change during this period be $d_t \equiv P_t - P_{t-1}$, and the return (or relative price change) be $r_t \equiv (P_t - P_{t-1}) / P_{t-1}$. Then we have: (a) $P(|r_t| > x) \sim Cx^{-\mu}$, as $x \to \infty$, where $\mu \approx 3$ and $C > 0$, and (b) $\mathrm{cor}(r_t, r_{t+h}) \approx 0$ for $h > 0$ (except perhaps for $h = 1$), but $\mathrm{cor}(|r_t|, |r_{t+h}|) > 0$ over a long range of lags $h$. FIG. 1 shows the two regularities for the NYSE daily index.

FIG. 1. NYSE composite daily index: (a) Price; (b) Return (in percentage); (c) Tail distribution of absolute return in log-log scale, showing a clear linear decay, and a least-square fit for values larger than 2%, with a slope of three (in absolute value); (d) The autocorrelation function of return is nearly zero at all lags, while that of absolute return is nonzero over a long range of lags (volatility clustering).

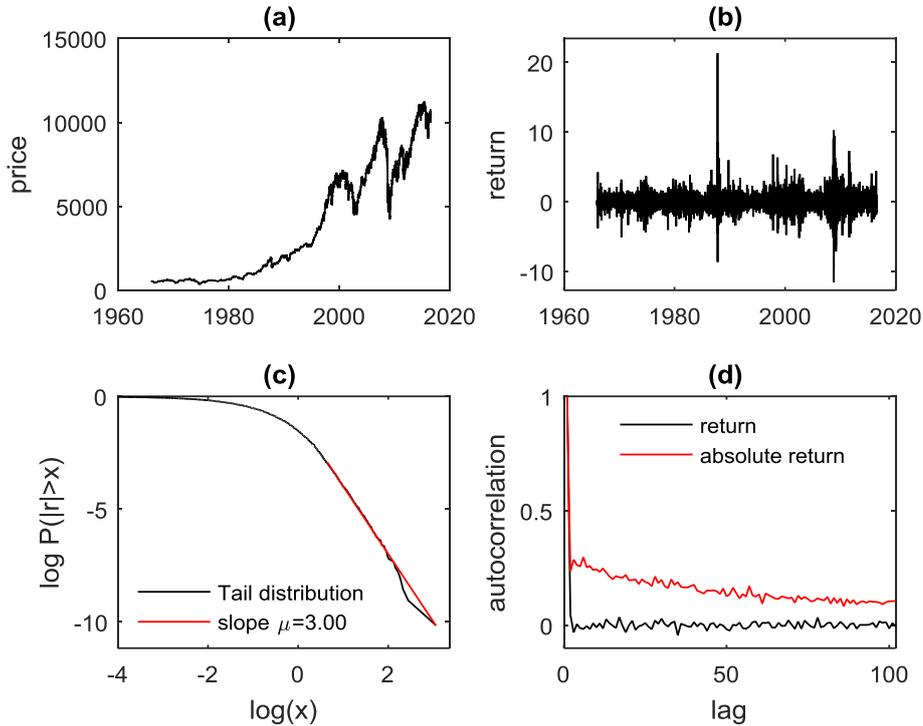

Many models have been proposed to account for these regularities, notably agent-based models, which often invoke relatively sophisticated mechanisms [1, 2]. This literature has played an important preliminary role, at least by raising the important questions and by uncovering the central concepts and mechanisms (feedback, heterogeneity, trend following, etc.). But it is time we seek the essence of these phenomena, by reducing the important concepts and mechanisms to their simplest formulation.

In a previous paper we show that speculation and trend-following are the essence of the power law of financial price changes, as they imply this regularity almost by definition: the return follows a random-coefficient autoregressive process in a speculative market, as long as the speculators' use past price changes to predict the future price change, making their expectations endogenous [3]. Thus the return has a power law by Kesten's theorem [4-6, 3]. But this model cannot explain volatility clustering, as another theorem by Basrak at al. implies [7-9, 6]. This paper extends this model by including, as is usual, a second class of agents, value-investors (or 'fundamentalists'), who attach an (intrinsic) value to the asset and buy it when they think it is underpriced (or sell it, otherwise). When the agents' expectations are exogenously driven, in the sense of being entirely shaped by exogenous news, then they can be simply modelled as following a random walk. This basic random walk model of expectations explains volatility clustering in a natural way.

## 2. The model

The market is populated by *speculators*, who buy an asset for a purely speculative reason, that is, when they expect its price to rise (or sell it, otherwise), and *value-investors* (or 'investors' for short), who attach a value to the asset and buy it when they think it is worth more than it is currently priced (or sell it otherwise). Formally, we assume the demand of the $i$th speculator is

$$x_{it} = \alpha d_{it}, \tag{1}$$

where $d_{it}$ denotes the price change that the speculator expects to occur in period $t$, and $\alpha > 0$. We assume on the other hand that the demand of the $j$th investor is

$$x_{jt} = \gamma (V_{jt} - P_t), \tag{2}$$

where $V_{jt}$ denotes the value that the investor expects the security to be worth in the end of period $t$, and $\gamma > 0$. As usual, supply is formally identified as a negative demand. We let $N_t$ and $M_t$ be respectively the numbers of speculators and investors who desire to buy or sell the security during period $t$. The overall excess demand can be written as

$$x_t = \alpha N_t \bar{d}_t + \gamma M_t (\bar{V}_t - P_t), \tag{3}$$

where $\bar{d}_t \equiv N_t^{-1} \sum_i d_{it}$ and $\bar{V}_t \equiv M_t^{-1} \sum_j V_{jt}$, namely the expected price change that speculators expect to occur on average, and the value that investors on average expect the asset to be worth, which we shall refer to simply as 'the value' of the security.

Finally, we assume that the price adjusts linearly to the overall supply-demand imbalance:

$$d_t = \beta x_t, \tag{4}$$

where $\beta > 0$. The dynamics of price is then given by $P_t \equiv P_0 + \sum_{k=1}^{t} d_k$ and (3) and (4) combined, which can be written as

$$d_t = a_t \bar{d}_t + b_t (\bar{V}_t - P_t), \tag{5}$$

where

$$a_t \equiv \alpha \beta N_t, b_t \equiv \gamma \beta M_t. \tag{6}$$

So the dynamics of the price depends crucially on how expectations are formed. We distinguish two cases: endogenous versus exogenous expectations. We say that expectations are endogenous if they depends on information generated by the market dynamics itself, and we say they are exogenous, otherwise, that is, if they are determined by exogenous information.

## 2.1 Trend following and power law

A common example of endogenous expectations is trend following (or chart reading). Assume a purely speculative market, that is, $M_t = 0$, so that $d_t = a_t \bar{d}_t$, and assume the speculators' predictions of the future price change are functions of past price changes $(d_{t-1}, d_{t-2}, ...)$. Assume, to simplify, that these functions are linear, that is, assume $d_{it} = \sum_{k=1}^{K} \omega_{ikt} d_{t-k} + \varepsilon_{it}$, where $K$ corresponds to the furthest past that speculators overall consider relevant, $\omega_{jkt}$ are coefficients, and $\varepsilon_{jt}$ capture other influences on expectations. Then the price change follows a random-coefficient autoregressive process:

$$d_t = a_t \sum_{k=1}^{K} \bar{\omega}_{kt} d_{t-k} + e_t, \tag{7}$$

where $\bar{\omega}_{kt} \equiv N_t^{-1} \sum_{i=1}^{N_t} \omega_{ikt}$ and $e_t = \alpha \beta \sum_j \varepsilon_{jt}$. We know from Kesten theorem that under general conditions, $d_t$ converges in distribution to a unique strictly stationary and ergodic process, for any initial value $d_0$. Moreover, under also general conditions, the limiting distribution is a power law, that is, there is some $\mu > 0$ such that $P(d_t > x) \sim Cx^{-\mu}$, where $C > 0$. Assuming $d_t$ is independent of $P_t$, which is reasonable on short periods, this implies the same power law for return, since, if $f(P,t)$ is the density function of the price,

$$P(r_t > x) \sim \int P(d_t > xP) f(P,t) dP \sim CE(P_t^{-\mu}) x^{-\mu}.$$

But again this model cannot explain volatility clustering, by a theorem by Basrak et al. [7, 8]. This theorem implies that for any measurable function $g$, $\text{cov}[g(d_t), g(d_{t+h})]$, if well-defined, decays exponentially with $h$. So both $\text{cov}(|d_t|, |d_{t+h}|)$ and $\text{cov}(d_t^2, d_{t+h}^2)$, for instance, decay exponentially with $h$ (if they are well-defined). To account for clustered volatility we should therefore model expectations differently.

## 2.2 Exogenous news and clustered volatility

Now let the agents' views be entirely driven by exogenous news. Formally, let $E_t$ and $F_t$ be the events associated with the arrival in period $t$ of news relevant to speculators and investors, respectively; let the probabilities of these events be $p(E_t) = \tau$ and $p(F_t) = \tau'$. We assume that, when they arrive, the news change the agents' views by $\varepsilon_t$ and $\nu_t$, respectively, which we assume to be normally distributed with zero mean and standard deviations $\sigma_\varepsilon$ and $\sigma_\nu$. In sum, we assume,

$$\bar{d}_t = \bar{d}_{t-1} + \varepsilon_t 1(E_t), \tag{8}$$

$$\overline{V}_t = \overline{V}_{t-1} + \nu_t 1(F_t), \tag{9}$$

where $1(E_t)$ and $1(F_t)$ are the indicator functions of $E_t$ and $F_t$. The implication of this simple random walk model of expectations is that the agents' views reflect all the information that came to them: $\overline{d}_t = \sum_{k=1}^t \varepsilon_k 1(E_k)$ and $\overline{V}_t = \overline{V}_0 + \sum_{k=1}^t \nu_k 1(F_k)$, where we let $\overline{d}_0 = 0$. The price dynamics is now given by $P_t \equiv P_0 + \sum_{k=1}^t d_k$ and

$$d_t = a_t \sum_{k=1}^t \varepsilon_k 1(E_k) + b_t [\overline{V}_0 + \sum_{k=1}^t \nu_k 1(F_k) - P_t]. \tag{10}$$

This simple model exhibits volatility clustering, as FIG. 2 shows, which is a simulation of (10) using exponential distributions for $a_t$ and $b_t$ with respective means 0.1 and 0.3, $\tau = 1$, $\tau' = 0.1$, $\sigma_\varepsilon = 0.1$, $\sigma_\nu = 1$, and $P_0 = \overline{V}_0 = 100$.

FIG. 2. Price changes when expectations are driven by exogenous news. (a) Price and value (for clarity, only the 500 first data are shown); (b) Return (in percent); (c) Tail distribution of absolute return in log-log plot and a least-square fit for values larger than 1.5%; (d) Autocorrelation function of return and absolute return.

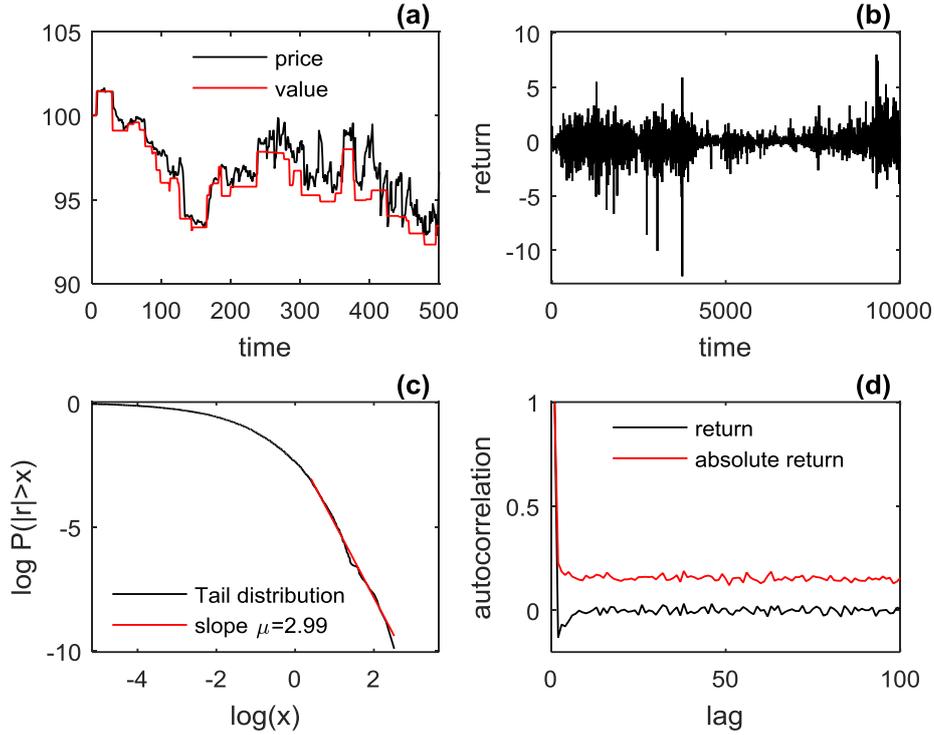

## 3. Discussion

The volatility clustering is generic in this model, in that it holds for different distributions of $a_t$ and $b_t$ and for a broad range of values of the parameters (as we have checked). Most parameters play only a quantitative role: $\sigma_\varepsilon, \sigma_\nu, \tau$ and $\tau'$ are chosen in this simulation merely to 'calibrate' the model with the real daily data displayed in FIG.1, namely to have a standard deviation of return around one percent. The distribution of return in this simulation is close to a power law with cubic exponent, as is the case in the empirical data [compare FIG.1(c) with FIG. 2 (c)]. (The cutoff $x_{min}$ beyond which the linear fit of the tail distribution is applied, which is 1.5 in this simulation, is determined throughout by an algorithm developed in paper [10].) But the underlying process being clearly non-stationary, given the random walks (8) and (9), the tail of return cannot be an exact and stationary power law as would be the case in a purely speculative market with trend-following speculators, where the process converges in distribution to a power law by a theorem [3]. But the distribution is always fat-tailed (on average over 10 realizations of the process with the above parameters, the kurtosis is 27, as in the real data of FIG.1). Also, the properties of $a_t$ and $b_t$ (namely their means in this particular choice of the exponential distribution) play only a quantitative role for the curvature of the distribution.

The key to volatility clustering in this model is the fact that the news have a permanent impact on the agents' views, as implied by the random walk model of expectations. This is a nontrivial result, however, as this persistence of the impact of news on expectations is reflected only in the amplitude of return, and not in its sign; indeed the return itself is essentially serially uncorrelated [FIG. 2 (d)].

But, actually, it is the speculators' expectations that matter for volatility clustering, as this regularity holds even when $\bar{V}_t$ is constant ($\tau' = 0$), as FIG. 3 shows.

FIG. 3. The same process as in FIG. 1 but keeping the asset's value constant. (a) Price and value; (b) Return (in percentage); (c) Tail distribution of absolute return in log-log plot and a least-square fit for its values larger than 1.5%; (d) Autocorrelation function of return and absolute return. The price changes are essentially the same as in FIG. 1.

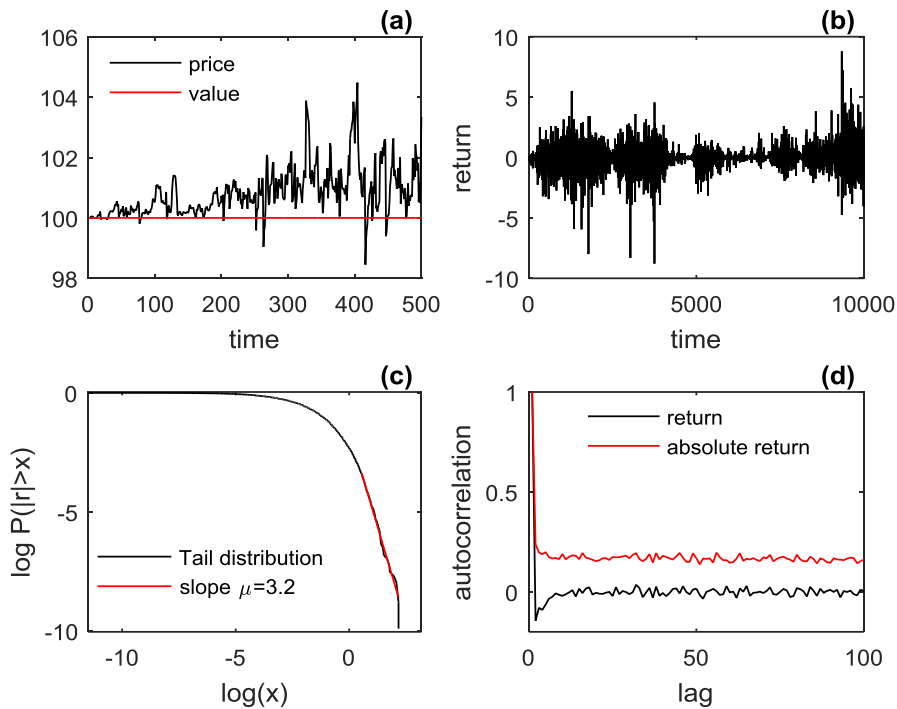

We choose $\tau = 1$ so that volatility clustering is not mistaken for a 'regime shift' in the return process, caused by a rare but consequential, regime-shifting, news (corresponding to a small $\tau$), a conclusion that the 'occasional structural breaks' literature may have prepared one to jump to [11-13]. For instance, Granger and Hyung suggest to model absolute return directly as following a random walk with 'occasional structural breaks'. But (apart from its ad hoc nature), this model has the drawback of implying a divergent absolute return, an explosive volatility, as the authors themselves point out [13].

In sum, this paper suggests a simple explanation for excess and clustered volatility in financial markets. Excess volatility for an asset means that its price changes are too high given the underlying fundamentals: the price will fluctuate even when the underlying value of the asset is constant. Clustered volatility simply reflects, in this model, the flow of exogenous news affecting the expectations of speculators.